\documentclass[aps,prb,twocolumn,groupedaddress,floatfix]{revtex4}
\usepackage{graphics,graphicx}
\usepackage{color}
\usepackage{amssymb,amsmath}
\usepackage{bm}

\newcommand{\vq}{{\bm{q}}}

\newcommand{\be}{\begin{equation}}
\newcommand{\ee}{\end{equation}}
\newcommand{\bea}{\begin{eqnarray}}
\newcommand{\eea}{\end{eqnarray}}
\begin{document}
\unitlength = 1mm
%~~~~~~~~~~~~~~~~~~~~~~~~~~~~~~~~~~~~~~~~~~~~~~~~~
\title{Momentum dependence and nodes of the superconducting gap in iron-pnictides}
\author{A.~V.~Chubukov$^1$, M.~G.~Vavilov$^1$, and  A.~B.~Vorontsov$^2$ }
\affiliation{$^1$~Department of Physics,
             University of Wisconsin, Madison, Wisconsin 53706, USA\\
$^2$~Department of Physics, Montana State University, Bozeman, MT, 59717, USA}

\date{\today}

%~~~~~~~~~~~~~~~~~~~~~~~~~~~~~~~~~~~~~~~~~~~~~~~~~~~~~~~~~~~~~~~~~~~~~~~~~~~~~
\begin{abstract}
We discuss the structure of the superconducting gap in iron pnictides.
In the itinerant electron picture, gaps with or without nodes have the
extended $s-$wave ($s^+$) symmetry and emerge within the same
pairing mechanism, determined by the interplay between
intra-pocket repulsion and inter-band pair hopping.
If the pair hopping is stronger, the system
develops an $s^+$ gap without nodes.
In the opposite case the superconductivity
 is governed by of the momentum-dependent part of the pair-hopping,
and an $s^+$ gap shows nodes on electron Fermi surfaces.
We argue that the gap without/with nodes emerges in
systems with a stronger/weaker tendency towards a spin order.
\end{abstract}
%~~~~~~~~~~~~~~~~~~~~~~~~~~~~~~~~~~~~~~~~~~~~~~~~~~~~~~~~~~~~~~~~~~~~~~~~~~~~~
\maketitle
%~~~~~~~~~~~~~~~~~~~~~~~~~~~~~~~~~~~~~~~~~~~~~~~~~~~~~~~~~~~~~~~~~~~~~~~~~~~~~

\begin{figure}
\centerline{\includegraphics[width=0.98\linewidth]{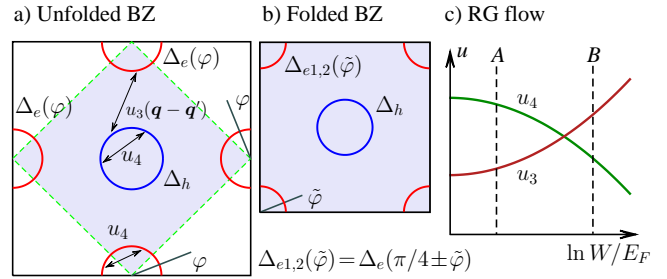}}
\caption{(Color online).
(a) Hole (center) and electron (edges) Fermi surfaces in the unfolded BZ
and interactions $u_4$ (intra-band repulsion) and  $u_3(\vq -\vq')$ (inter-band hopping).
The $s^+$ SC gap is nearly uniform on the hole FS but
may have nodes along the dashed lines near the electron FS.
(b) The folded BZ.
Each corner now hosts two electron FSs with gaps $\Delta_{e1,2}({\tilde \phi})$.
(c) Schematic RG flow for $u_4$
and the momentum-independent part of $u_3(\vq -\vq')$
up to energies $ \sim E_F$.
The couplings at this scale determine the pairing.
If renormalized $u_3 > u_4$ (line B),
the SC pairing occurs even for $u_3(\vq -\vq')={\rm const}$,
while for $u_3<u_4$ (line A)
the pairing is induced by the momentum-dependent part in $u_3(\vq -\vq')$,
and $\Delta_{e}(\varphi)$ has nodes.}
\label{fig:1}
\end{figure}

{\it Introduction.} The structure
of the superconducting (SC)
 gap is one of the most controversial topics in the rapidly growing
field of iron-based pnictide superconductors.
Electronic band configuration of the pnictides assumes two hole pockets
centered at the $\Gamma$ point $\vq = (0,0)$ and two electron pockets
centered at $(0,\pi)$ and $(\pi,0)$ in the unfolded
Brillouin zone (BZ) to which we will refer in this paper, Fig.~\ref{fig:1}a.
Multiple Fermi surfaces (FSs)
create a number of different possibilities for the
 gap structures.\cite{gorkov}

Majority of theoretical works
  predict gaps of one sign on the hole FSs,
and of another sign, on average, on the electron FSs.
\cite{mazin,others,bang,bernevig_2,scal,d_h_lee,chubukov08,k_a} We
will refer to such gap structure as an extended $s^+$ state. There
is no consensus, however, on whether the $s^+$ order parameter has
nodes on electronic FS. Gaps without nodes have been found in the
2-band and 5-band itinerant
models\cite{bang,d_h_lee,chubukov08,others,k_a} and in the localized
spin models.\cite{bernevig_2} On the other hand Graser {\it et al}
found\cite{scal} an $s^+$ state with nodes in the 5-orbital Hubbard
model.

Previous results for the gap  structure
have been obtained
numerically, and there is a call for a simple analytical analysis
of the pairing mechanisms and the gap.
 In this paper, we show that
 different results for the gap structure
are not fundamentally conflicting,
and that both nodal and node-less $s^+$ gaps
 may emerge in the same pairing scenario.

Electron-phonon interaction is
 weak in Fe-pnictides~\cite{phonons} and
the pairing likely has an electronic origin.
In the weak-coupling approach, the most natural
candidate for a pairing glue is the
hopping of electron pairs between hole and electron pockets
($u_3$ term in Fig.~\ref{fig:1}a).
For small pockets, a repulsive pair hopping gives rise to an $s^+$ state with
the sign change of the gap between hole and electron FSs, but without nodes,
due to small, in parameter $E_F/W$, angular variations of the gaps along the FSs,
where $W$ is the fermionic bandwidth.
This simple scenario is, however, incomplete as there also exists a repulsive
interaction within each pocket ($u_4$ in Fig.~\ref{fig:1}).
This interaction does not cancel out from the equation for the $s^+$ gap,
and we show the $s^+$ state without nodes is only possible when at low energies
(smaller than $E_F$) the  pair-hopping term exceeds the intra-pocket repulsion.

%%%%%%%%%%%%%%%%%%%%%%%%%%%%%%%%%%%%%%%%%%%%%%%%%%%%%%%%%%%%%%%%%%%%%%%%%%%%%%
%%%%%%%%%%%%%%%%%%%%%%%%%%%%%%%%%%%%%%%%%%%%%%%%%%%%%%%%%%%%%%%%%%%%%%%%%%%%%%

The restriction to energies smaller than $E_F$ is essential here
because bare intra-pocket repulsion is very likely larger than the
pair hopping term,
and the latter may win only if
the two interactions are renormalized  by high-energy
fermions and flow from their bare values at energies of order $W$
to new values at energies $\sim E_F$, Fig. \ref{fig:1}c.
Because of near-nesting between at least
one hole and one electron FSs the renormalizations
in particle-particle and particle-hole channels are equally important
in between $E_F$ and $W$,
 and the analysis of the appropriate parquet renormalization group (RG)
shows~\cite{chubukov08} that the couplings' flow is controlled by
a density-density inter-pocket interaction
which determines the system's tendency towards a spin density wave (SDW) order.
When this tendency is strong, pair-hopping coupling gets pushed up.
The intra-pocket repulsion meanwhile decreases under RG
as one should expect for a repulsive pairing interaction. As a result,
when the system has a strong tendency towards a SDW order,
the pair-hopping term  becomes larger than the intra-pocket repulsion
somewhere between $W$ and $E_F$, and the system develops an $s^+$ gap without nodes.

We discuss here what happens for weaker
renormalization of the coupling constants,
when the pair-hopping interaction remains smaller than the intra-pocket
repulsion at energies of order $E_F$.
Our key message is that, despite strong repulsion,
the system still develops a pairing instability due to
momentum-dependent part of
the pair-hopping term. This instability leads to the
$s^+$ gap with nodes on the two electron FS
and occurs in the clean limit {\it at an arbitrary weak pair-hopping}.
The reason why the pairing occurs despite strong
repulsion is quite generic and is related to the fact that the gap
that oscillates along the electron FS is insensitive to the
intra-pocket repulsion.\cite{scal_1}
The gap at the hole FS is still affected by
the intra-pocket repulsion, but it turns out
that the elimination of the repulsion from just the electron FS
is a sufficient condition for the
 SC instability, still driven by the pair-hopping
term between electron and hole FSs.

{\it The gap equation.}
We label fermions near different FSs as $f$ (hole) and $c$ (electron).
For simplicity, we consider the case when the
gaps $\Delta_h (\vq)$ are equal on the two hole FSs,
and the gaps $\Delta_e (\vq)$ along the two electron FSs transform
 into each other under rotations by $\pi/2$,
and reflections in all symmetry planes.
The extension to the case of non-equal gaps on two hole FSs is straightforward.
For a model of itinerant electrons, we consider the interactions
within each pocket
$u_4 (\vq-\vq') (c^\dagger_\vq c^\dagger_{-\vq} c_{\vq'} c_{-\vq'} +
f^\dagger_\vq f^\dagger_{-\vq} f_{\vq'} f_{-\vq'})$ ,
and between the pockets
$u_3 (\vq-\vq') c^\dagger_\vq c^\dagger_{-\vq} f_{\vq'} f_{-\vq'} + h.c$
 ~(we  borrowed the notations from Ref.~\onlinecite{chubukov08}).
Functions $u_3 (\vq-\vq')$ and $u_4 (\vq-\vq')$  have projections
onto one-dimensional $A_{1g}$ ($s$-wave) and $B_{1g}$ ($d-$wave) and
two-dimensional $E$ ($p-$wave) representations
 of the crystal symmetry.
 We focus on the $A_{1g}$ component
$u_3^{A_{1g}} (\vq,\vq')\propto Y_{A_{1g}}(\vq)Y_{A_{1g}}(\vq')$ with
$Y_{A_{1g}}(\vq) = 1 + b(\cos q_x + \cos q_y)$ ($b$ is a constant).
For small hole and electron pockets,
 $Y_{A_{1g}}(\vq) \approx 1+2b$ on the hole FSs and
$Y_{A_{1g}}(\vq)=1-(b k_F^2/2) \cos 2\varphi$ on the electrons FSs,
 see Fig.~\ref{fig:1}a.
 Then $u_4 (\vq-\vq') \approx u_4$, and
$u_3^{A_{1g}} (\vq,\vq') \approx u_3+\sqrt{2}\tilde u_3\cos 2\varphi$
($\varphi$ corresponds to $\vq$ or $\vq'$, whichever is on the electron FS).
 We also studied a $B_{1g}$ gap, but found that it emerges at a lower $T_c$~\cite{footnote}.

We assume that the couplings  $u_3$, $\tilde u_3$ and $u_4$, are already
renormalized by fermions with energies between $W$ and $E_F$, see
Fig.~\ref{fig:1}c, and consider the system behavior below $E_F$, when the
pairing channel is decoupled from the SDW channel,
and the SC instability problem can be treated within the BCS approximation.
The set of the coupled BCS equations for $\Delta_h (\vq)$ and $\Delta_e (\vq)$ is obtained in the standard manner.  On the hole FSs
 $\Delta_h (\vq)$ is approximately a constant  $\Delta_h$,
while on electron FSs
 $\Delta_e (\vq) = \Delta_e  + {\bar \Delta}_e (\cos q_x + \cos q_y)
\approx \Delta_e + \tilde\Delta_e \sqrt{2} \cos 2 \varphi$
with ${\tilde \Delta}_e = -k_F^2{\bar \Delta}_e/2\sqrt{2}$.
 The coupled linearized equations  for three gaps $\Delta_h$, $\Delta_e$, and
${\tilde \Delta}_e$ are
  (we take all FSs as cylindrical)
\begin{subequations}\label{eq:1}
\bea
\Delta_h &= & - u_4 L \Delta_h   - u_3 L \Delta_e  -  {\tilde
u}_3 L
{\tilde \Delta}_e;  \label{eq:1a}
\\
\Delta_e &= & -u_4 L \Delta_e  - u_3 L \Delta_h;\\
\label{eq:1b}
{\tilde\Delta}_e &= & -  {\tilde u}_3 L \Delta_h,
\label{eq:1c}
\eea
\end{subequations}
where
$L = \ln (1.13 \Lambda/T_c)$,
$\Lambda \sim E_{\rm F}$ is the upper energy cutoff.
  From Eqs.~\ref{eq:1} we obtain the equation for the critical temperature
$T_c$ in the form
\be
(1 + u_4 L)^2 - u^2_3 L^2 = {\tilde u}^2_3  (1 + u_4 L) L^2.
\label{3}
\ee
It is invariant with respect to sign change of
${\tilde u}_3$ and $u_3$.

The gap that emerges at $T_c$ has generally  nonzero $\Delta_h$,
$\Delta_e$, and  ${\tilde \Delta}_e$. Their ratios are
\be
\frac{{\tilde \Delta}_e}{\Delta_e} =
\frac{{\tilde u}_3}{u_3}(1+u_4 L);\quad
\frac{\Delta_e}{\Delta_h} = - \frac{u_3 L}{(1 + u_4 L)} \,,
\label{6}
\ee
where $L$ is a solution of Eq.~(\ref{3}).
 Note that  $\Delta_e$ and $\Delta_h$ have opposite signs if $u_3>0$.
When the ratio ${\tilde \Delta}_e/\Delta_e <1/\sqrt{2}$,
the $s^+$ state has no nodes, otherwise there are nodes on the electron FS.

Equation (\ref{3}) is a cubic equation in $L$ and can be
analyzed for arbitrary $\delta = {\tilde u}_3/u_3$ and $\gamma = u_4/u_3$.
For $\delta =0$, it
 gives $L = L_0 = 1/(u_3-u_4) = 1/(u_3 (1 -\gamma))$,
and a solution for finite $T_c$ exists only
when   $\gamma <1$. The gap satisfies
$\Delta_e = - \Delta_h$ and ${\tilde \Delta}_e =0$.

\begin{figure}
\centerline{\includegraphics[width=0.9\linewidth]{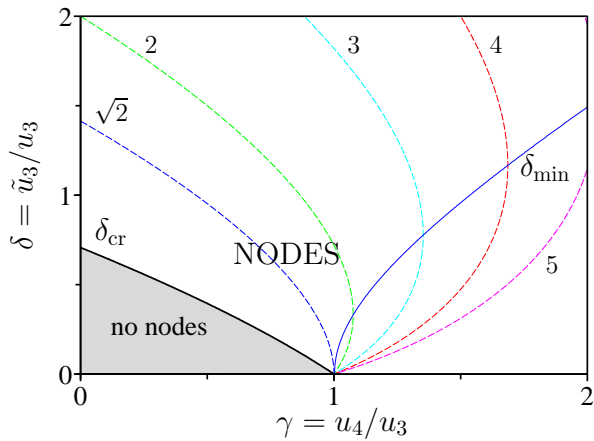}}
\caption{(Color online).
The phase diagram in dimensionless interactions %AV - for different -
$\gamma = u_4/u_3$ and $\delta = {\tilde u}_3/u_3$. The nodeless
$s^+$ state exists in the lower left corner below the line
$\delta_{\rm cr}(\gamma)$, when intra-band repulsion $u_4$ and
momentum-dependent inter-band hopping $\tilde{u}_3$ are small.
Dashed lines show the contours of fixed ratio $\tilde
\Delta_e/\Delta_e=\sqrt{2},\ 2,\ 3,\ 4,\ 5$. The narrow solid line
shows the position of a minimum of $\tilde \Delta_e/\Delta_e$ in the
region where the gap has nodes. Below this line, the gap develops
nodes primarily to reduce the effect of  intra-pocket repulsion.}
\label{fig:2}
\end{figure}

Consider now $\delta \neq 0$.
A naive expectation would be a smooth
monotonic
evolution of the gap with increasing $\delta$.
One indeed finds a smooth evolution, but only for $\gamma <1$,
when superconductivity exists even at $\delta =0$.
For such $\gamma$, $L$ gradually decreases with $\delta$, and
the  oscillating component ${\tilde \Delta}_e$ continuously  increases.
The nodes on the electron FS appear above some critical $\delta_{cr}(\gamma)$,
see Fig.~\ref{fig:2}. We found
$\delta_{cr}(\gamma \ll 1) \approx 1/\sqrt{2} -\gamma/ \sqrt{3}$
 and $\delta_{cr}(\gamma ) \approx \sqrt{8/9} (1-\gamma) $ for $1-\gamma\ll 1$.

For $\gamma \geq 1$, we found a new behavior: the
superconductivity develops for any $\delta \neq 0$.
Indeed, the r.h.s of Eq.~(\ref{3}) scales as $u^2_3 (\gamma^2 -1) L^2$
 at $L\gg 1$,
 while the l.h.s. scales as $u^3_3 \gamma \delta^2 L^3$.
Comparing the two, we find $L = (\gamma^2 -1)/(u_3 \delta^2\gamma)$, i.e.
$T_c$ remains finite
 (but exponentially small for $\delta\to 0$):
 \begin{equation}
T_c= 1.13 \Lambda \exp\left(-\frac{\gamma^2-1}{u_3 \gamma \delta^2}\right).
 \label{eq:Tc}
\end{equation}
The oscillating  component of the gap ${\tilde \Delta}_e$
now dominates and exceeds both $\Delta_h$ and $\Delta_e$ by $1/\delta$.
The reason why $T_c$ is non-zero even when the intra-pocket repulsion is
stronger than the pair hopping (i.e., when $\gamma \gtrsim 1$)
is the absence of the $u_4 {\tilde \Delta}_e$ term in Eq.~(\ref{eq:1c})
because the angular integral of ${\tilde \Delta}_e(\varphi)$
vanishes along the electron FS. In other words,
oscillations of the gap along the electron FS allow a
 system to avoid strong intra-pocket repulsion and develop SC order.

We see
that there is a qualitative change in the system behavior near  $\gamma \approx 1$ and at
$\delta \ll 1$. In this region
${\tilde \Delta}_e/\Delta_e =[\sqrt{2 \delta^2 + (1-\gamma)^2} + (\gamma-1)]/\delta$
is nearly  discontinuous, evolving from ${\cal O}(\delta)$ at $\gamma <1$
to $\sqrt{2}$ at $\gamma =1$, and to
${\cal O}(1/\delta)$ at $\gamma >1$. The ratio $\Delta_e/\Delta_h \approx -1$
at $\delta\ll 1$ and $\gamma \leq 1$, and decreases as $\Delta_e/\Delta_h
\approx -1/\gamma$ for $\gamma \gg 1$.

%%%%%%%%%%%%%%%%%%%%%%%%%%%%%%%%%%%%%%%%%%%%%%%%%%%%%%%%%%%%%%%%%%%%%%%%%%%%%%%%%%%%
%%%%%%%%%%%%%%%%%%%%%%%%%%%%%%%%%%%%%%%%%%%%%%%%%%%%%%%%%%%%%%%%%%%%%%%%%%%%%%%

On a more careful look, we find two sub-regimes for $\gamma >1$, when the gap
has nodes. One is the regime of small $\delta$, where the gap is fully adjusted
to minimize the effect of repulsive $u_4$. In this regime, ${\tilde \Delta}_e
\gg \Delta_e, \Delta_h$, i.e., the gap is essentially $\cos 2 \varphi$ on
electron FS and much smaller in magnitude on the hole FS. The other is the
regime of large $\delta$, where the ratio ${\tilde \Delta}_e/\Delta_e$ is again
large, but not because the system tends to avoid $u_4$ but rather because the
momentum-dependent part of the effective pair-hopping becomes the dominant
interaction.
In between, ${\tilde \Delta}_e/\Delta_e$ passes through minimum, at
$\delta = \delta_{\rm min}(\gamma)$.
Near $\gamma =1$, $\delta_{\rm min}(\gamma) \approx \sqrt{\gamma-1}$,
while for large $\gamma$, $\delta_{\rm min}(\gamma) \approx (\sqrt{3}/2) \gamma$.
The ratio ${\tilde \Delta}_e/\Delta_h$ monotonically
decreases with increasing $\delta$ and is not affected by the change of the
physics upon crossing of $\delta_{\rm min} (\gamma)$.

The generic phase diagram in the $(\gamma,\delta)$ plane
is presented in Fig.~\ref{fig:2}. The bolder solid line is
$\delta_{\rm cr}(\gamma)$ above which  $s^+$ gap acquires nodes, and the narrower solid line is
$\delta_{\rm min} (\gamma)$,  where the ratio $\tilde\Delta_e/\Delta_e$ is at minimum.

%%%%%%%%%%%%%%%%%%%%%%%%%%%%%%%%%%%%%%%%%%%%%%%%%%%%%%%%%%%%%%%%%%%%%%%%%%%%%
\begin{figure}
\centerline{\includegraphics[width=0.95\linewidth]{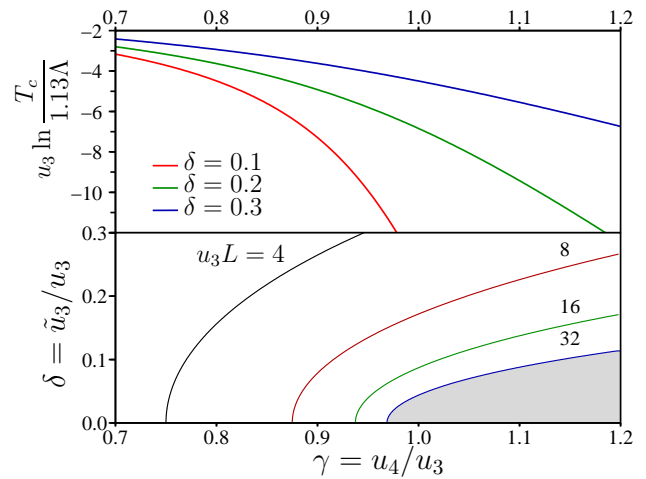}}
\caption{(Color online). Top: dependence of the critical
temperature, $T_c$, in a clean system, on the parameter
$\gamma=u_4/u_3$ for three values of $\delta=0.1;\ 0.2;\  0.3$. For
small $\delta$, the critical temperature drops fast as $\gamma$
approaches  $\gamma=1$. For larger $\delta$,
 the critical temperature remains relatively large even above $\gamma=1$.
Bottom:
contours of constant $u_3 L = u_3 \ln 1.13 \Lambda/T_c =4,\ 8,\ 16, \ 32$
in the ($\gamma, \delta$) plane.}
\label{fig:3}
\end{figure}
%%%%%%%%%%%%%%%%%%%%%%%%%%%%%%%%%%%%%%%%%%%%%%%%%%%%%%%%%%%%%%%%%%%%%%%%%%%%%%%

{\it Role of spin fluctuations.}~~~~ Above we considered only a
direct four-fermion pairing interaction (terms to first order in
$u_{3,4}$). The antisymmetrized pairing vertex created from this
interaction has equal charge and spin components. Beyond first
order, such vertex contain non-equal spin and charge components, and
the spin component describes the interaction mediated by collective
spin fluctuations.\cite{chubukov08} Such interaction is enhanced
near an SDW instability and, like $u_3$, is attractive for a
plus-minus $s^+$ gap. Accordingly, it extends the boundary  of an
$s^+$ state without
 nodes towards  larger $\gamma$.\cite{schmalian}  Still, unless the system is right at the SDW
instability point, a SC with a plus-minus gap is not realized for sufficiently
large $u_4$, and, as before, one needs a momentum-dependent part of the
pair-hopping (or of a spin-fluctuation exchange) to get $T_c \ne 0$. This, in
turn, gives rise to a gap with nodes on the electron FS.

{\it Role of impurities.}
In the presence of potential impurities, the scattering of electrons results in
mixing of electron states with different momenta. This mixing does not affect
superconductors with isotropic order parameter.
The plus-minus $s^+$ SC state at $\gamma <1$ and $\delta \ll 1$ is suppressed
by inter-pocket impurity scattering $\Gamma_\pi$,\cite{chubukov08,mazin_imp,anton_2} but
is not sensitive to a much stronger intra-pocket impurity scattering $\Gamma_0$.
We now show that $\Gamma_0$ strongly reduces the critical temperature $T_c(\Gamma_0)$
at  $\gamma \gtrsim 1$, when superconductivity develops through an
anisotropic component of the order parameter on
the electron Fermi surfaces.
To evaluate the effect of $\Gamma_0$ on $T_c$,
we derived the self-consistency equations within the Born approximation.
We found that the
linearized gap equations can be presented in the form of Eqs.~(\ref{eq:1}),
but with Eq.~(\ref{eq:1a}) for the hole FS gap replaced by
\begin{equation}
\begin{split}
\Delta_h & = - u_4 L \Delta_h   - u_3 L \Delta_e  -  {\tilde u}_3 {\cal L}(\Gamma_0)
{\tilde \Delta}_e, \nonumber \\
   {\cal L}(\Gamma_0) & =
\ln\frac{1.13 \Lambda}{T_c(\Gamma_0)} + \psi\left(\frac{1}{2}\right)
-\psi\left(\frac{1}{2} +\frac{\Gamma_0}{2\pi T_c(\Gamma_0)}\right) ,
\end{split}
\nonumber
\end{equation}
 and $\psi(x)$ is the digamma function.
The equation for $T_c(\Gamma_0)$ becomes
\be
(1 + u_4 L)^2 - u_3^2 L^2 = {\tilde u}_3^2  (1 + u_4 L) L  {\cal L}(\Gamma_0) \,.
\label{3prime}
\ee
Both the l.h.s  and  the r.h.s of Eq.~(\ref{3prime}) now scale as
$L^2$ at  $T_c \to 0$ because
${\cal L}(\Gamma_0)$ tends to a constant at $T_c=0$.
As a result, the solution of Eq.~(\ref{3prime}) does
 not exist for sufficiently small values of  ${\tilde u}_3^2$,
 i.e., intra-pocket impurity scattering is pair-breaking for
 superconductivity driven by momentum-dependent pair hopping.
For $\gamma > 1$ and $\delta \ll 1$,
 $T_c (\Gamma_0)$ vanishes when
 $\Gamma_0= T_c/1.13$, where $T_c$ is the transition temperature in clean samples.

{\it Application to the pnictides.}
The electron and hole pockets in the pnictides are rather small in size, and it
is likely that $\delta \propto E_F/W$ is small.\cite{comm_a}
Our analysis shows that for small $\delta$, two situations are
possible.  When the RG renormalizations
 of $u_{3,4}$  are strong enough, due to a tendency to a
nesting-driven  SDW order, $\gamma = u_4/u_3$ is smaller than 1 at energies of
order $E_F$, and the system develops a plus-minus gap without nodes and
near-equal magnitudes along nested hole and electron FS. In this situation, the
leading instability of undoped system is towards an SDW order, so one should
expect that it
  becomes a superconductor only after spin order is destroyed by a finite
 doping which acts against nesting~\cite{vvc}.
 We believe this scenario should work for
 $FeAs$ $1111$ and $122$ materials which do show an SDW order at small doping
 and superconductivity at larger dopings.

If the tendency towards SDW order is less strong and $u_4 - u_3$ remains
positive down to $E_F$ (i.e., $\gamma >1$), then the pairing is predominantly
determined by the angle-dependent component of the pair-hopping term.
In this situation, $T_c$ is smaller, and the pairing gap has nodes on the
electron FS.
This is consistent with what is currently known about $FeP$ 1111 material
$LaOFeP$ which does not display a SDW order
and has a small $T_c \sim 5K$.

This may explain diverse reports of the measured temperature dependence of the
penetration length $\Delta \lambda (T)$
The exponential $T$ dependence of $\Delta \lambda (T)$  in
$SmFeAsO$~\cite{carrington_1} is consistent with the nodeless $A_{1g}$ gap;
linear in $T$ behavior in $LaOFeP$~\cite{carrington} is consistent with line nodes.
Measured $\Delta \lambda (T)$ in $K$ or $Co-$ doped  $BaFe_2As_2$ and in
$F-$doped $La(Nd)FeAsO$~\cite{proz} scales roughly as $T^2$. The latter behavior
is difficult to obtain from
the momentum-dependence of the $s^+$ gap alone,
and likely originated from the effect of impurities: either
inter-pocket impurity scattering $\Gamma_\pi$ in an $s^+$ superconductor
 without
nodes,\cite{anton_2} or intra-pocket scattering $\Gamma_0$ for an $s^+$ SC with
nodes.\cite{scal_new}

{\it Comparison with other theories.} Our results are consistent
with the numerical RG studies by Wang et al~\cite{d_h_lee} who found
an extended $s-$wave state, but with a substantial modulation of the
gap along the two electron FS.  These authors found that the
modulation increases with doping, what is in line with the idea that
the upturn renormalization of  $u_3$, driven by a tendency towards a
SDW order, decreases with doping. Our results and the reasoning are
also consistent with RPA-based studies of the 5-orbital Hubbard
model by Grasier et al.\cite{scal} They considered the case of
near-equal intra-orbital and inter-orbital Hubbard interactions,
which in our notations corresponds to $u_3 \approx u_4$.  They found
that the gaps along electron FS have nodes and $\Delta_e (\varphi) =
\Delta_e (1 + r \cos 2\varphi)$ with $r \approx 2.3-2.5$. We also
found the gap with nodes at $u_3 =u_4$, and our results are $r = 2$
at $\delta = 0^+$, and $r=2.3$ at $\delta =0.27$. We recall that a
small value of $\delta$ is expected for small sizes of the pockets
because $\delta \propto E_F/W$.  We re-iterate that the gap with no
nodes appears only when $u_3$ flows to a higher value than $u_4$
under the parquet RG, due to non-ladder renormalizations originating
from  the mixing with the SDW channel. Such renormalizations are not
included into the RPA formalism.

{\it Summary.}
 We showed here
 that the gap structure in iron-pnictides is sensitive to the interplay between
 intra-pocket repulsion and the pair hopping at energies ${\cal O}(E_F)$.  If the pair
 hopping is larger, the system develops a gap without nodes but with different
 signs of the gap on electron and hole FSs. If the intra-pocket repulsion is
 larger, the pairing is still possible,  but is now governed by the
 momentum-dependent part of the pair-hopping. Such pairing yields an extended
 $s-$wave  gap with nodes on the electronic FS, which allows the system to
 avoid a strong intra-pocket repulsion.  We argue that the extended $s-$wave
 channel always wins over a $d-$wave channel and that the gap with nodes is
 more likely for the systems with less developed tendency towards a SDW order.
 This nodal state is, however, affected by intra-pocket impurity scattering.

We acknowledge with thanks useful discussions with
 E. Abrahams, A. Carrington, I. Eremin,  K. Haule,
 P. Hirshfeld, G. Kotliar, D-H. Lee, J-X. Li, T. Maier,
 R. Prozorov, D. Scalapino, J. Schmalian, and D. Singh.
 The work  was supported by NSF-DMR 0604406 (A. Ch).

{\it Note added.} A day before this paper was submitted to arXiv, T.
Maier {\it et al} posted
 arXiv 0903.5216 with their analysis of the gap anisotropy,
 based on numerical analysis of a 5 band model. They named the
 momentum dependence of the interaction and the need to overcome
a repulsion within hole and electron pockets as the reasons for the gap anisotropy and the nodes. These are the same reasons that we found in our analytical study.


\begin{thebibliography}{99}
\bibitem{footnote}
The equation  for the $d$-wave $T_c$
 is given by Eq. (\protect\ref{3})  without $u_3$,
$1+u_4L=\tilde u_3^2 L^2$, and yields
a smaller $T_c$ (larger $L$) than
Eq.~(\protect\ref{3}) for the $A_{1g}$ gap with $u_3\neq 0$.
Attractive $B_{1g}$ channel with smaller $T_c$ than $T^{A_{1g}}_c$
is consistent with numerical RG and RPA-type studies.
\protect\cite{d_h_lee,scal}.
\bibitem{comm_a} The value of $\delta$ is also influenced by the fact that the
electron FS are actually not circular and different parts of the FS have
different orbital character, see e.g. D.J. Singh, Phys. Rev. {\bf 78}, 094511
(2008).
\end{thebibliography}
\end{document}